\documentclass[preprint,12pt]{elsarticle}
\usepackage{setspace}
\usepackage[]{natbib}
\usepackage{graphicx,amssymb,amsmath,times}
\usepackage{subfigure}
\usepackage{lscape}
\journal{Experimental Astronomy}
\begin{document}
\begin{frontmatter}
\title{An Artificial Intelligence based approach for constraining the redshift of blazars using $\gamma$--ray observations}
\author[label1,label2]{K K Singh\corref{corr}}
\ead{kksastro@barc.gov.in}
\author[label2]{V K Dhar} 
\author[label1]{P J Meintjes}  
\address[label1]{Physics Department, University of the Free State, Bloemfontein- 9300, South Africa}
\address[label2]{Astrophysical Sciences Division, Bhabha Atomic Research Center, Mumbai- 400 085, India}
%---------------------------------------Abstract----------------------------------------------------------
\begin{abstract}
In this paper, we discuss an artificial intelligence based approach to constrain the redshift of blazars using combined 
$\gamma$--ray observations from the \emph{Fermi} Large Area Telescope (LAT) and ground based atmospheric Cherenkov 
telescopes (ACTs) in GeV and \emph{sub} TeV energy regimes respectively. The spectral measurements in GeV and TeV 
energy bands show a redshift dependent spectral break in the $\gamma$--ray spectra of blazars. We use this observational 
feature of blazars to constrain their redshift. The observed spectral information of blazars with known redshifts reported 
in the \emph{Fermi} catalogs (3FGL and 1FHL) and TeV catalog are used to train an Artificial Neural Network (ANN) based 
algorithm. The training of the ANN methodology is optimized using \emph{Levenberg - Marquardt} algorithm with $\gamma$--ray 
spectral indices and redshifts of 35 well observed blazars as input and output parameters respectively. After training, 
we use only observed spectral indices in GeV and sub TeV regimes for 10 blazars as inputs to predict their redshifts. 
The comparison of predicted redshifts by the ANN with the known redshift suggests that both the values are consistent 
within $\sim$ 18$\%$ uncertainty. The method proposed in the present work would be helpful in future for constraining or 
predicting the redshifts of the blazars using only observational $\gamma$--ray spectral informations obtained from the 
\emph{Fermi}-LAT and current generation IACTs as well as from the next generation Cherenkov Telescope Array (CTA) with 
improved source statistics.
\end{abstract}
%---------------------------------------Keywords-------------------------------------------------------------
\end{frontmatter}

%---------------------------------------Section-1:Introduction------------------------------------------------
\section{Introduction}  
Blazars are radio-loud Active Galactic Nuclei (AGN) constituting a major population of the TeV $\gamma$--ray sources 
in the extragalactic Universe. These sources are characterized by a supermassive black hole surrounded by an accretion 
disk at the center of a host galaxy with generaly elliptical morphology and a relativistic jet pointing towards the observer 
at Earth \cite{Urry1995}. The orientation of the jet close to the line of sight of the observer leads to the relativistic 
beaming effects like superluminal motion and strong anisotropic radiation in the non-thermal emission from the outflowing 
plasma. The observed luminosity from such sources outshines their host galaxy and the non-thermal continuum emission from the 
jet can extend over the entire electromagnetic spectrum from radio to very high energy (VHE: E $>$ 100 GeV) $\gamma$--ray. 
The strong non-thermal emission characterizes the broad-band spectral energy distribution (SED) of blazars with two 
characteristic humps peaking at low and high energies (HE: E $>$ 100 MeV) respectively. The origin of the low energy hump from 
radio to X-rays through optical/UV is attributed to the synchrotron emission from the relativistic electrons gyrating in 
the magnetic field of the jet. The synchrotron origin of low energy component in the blazar SED is completely understood and has 
been observationally supported by the measurements of the high degree of linear polarization in radio and optical 
bands \cite{Lister2018,Blinov2016}. The physical mechanism for the second component from MeV-GeV to VHE or TeV $\gamma$--rays is not 
very clear and various models based on the leptonic and  hadronic processes have been proposed in the 
literature \cite{Tavecchio1998,Aharonian2000,Bottcher2013,Zech2017}.
\par
Based on the position of the synchrotron peak in the low energy hump of the SED, blazars are classified as low-synchrotron peaked (LSP),
intermediate-synchrotron peaked (ISP) and high-synchrotron peaked (HSP) sources \cite{Abdo2010a}. For LSP blazars, the rest frame 
synchrotron peak frequency is lower than 10$^{14}$ Hz and ISP sources have peak frequency in the range 10$^{14}$ Hz to 10$^{15}$ Hz. 
The synchrotron peak frequency for HSP blazars is generally observed to be more than 10$^{15}$ Hz and up to 10$^{18}$ Hz for specific 
sources. According to Meyer et al. (2011), HSP blazars and radio galaxies of FR I type have weak relativistic jets whereas LSP blazars 
and FR II radio galaxies belong to the strong jet population \cite{Meyer2011}. Blazars are also classified as BL Lacertae objects 
(BL Lacs) and Flat Spectrum Radio Quasars (FSRQs) on the basis of their properties in the optical band \cite{Padovani1995}. 
The optical radiation from blazars is considered to have contributions from both thermal and non-thermal emissions. 
The thermal emission originates from the accretion onto the supermassive black hole and from the host galaxy of the source. 
The non-thermal contribution comes from the relativistic jet. The optical spectra of FSRQs show strong emission lines from the 
thermal plasma contribution whereas no or weak emission lines are observed from the BL Lacs. This indicates that the host galaxy 
features are not clearly visible in the optical spectra of BL Lacs as compared to the FSRQs. The observed blazar sequence 
(anticorrelation between bolometric luminosity and synchrotron peak frequency) suggests that FSRQs are high power LSP blazars 
whereas BL Lacs are low power HSP sources \cite{Giommi2012}.
\par
The featureless optical spectra of BL Lacs render serious challenges in the measurement of their redshift (z) using optical 
spectroscopic methods. The redshift of cosmic sources like blazars is one of the important concepts of astrophysics and can 
be measured directly in observational cosmology. It plays a very important role in probing the evolution and structure formation 
in the Universe. In particular, the evolutionary properties of blazars is an open question in AGN astrophysics and cosmology. 
Apart from the cosmological evolution, the redshift of blazars is crucial for interpreting their multi-wavelength emission 
with different models and  to study the intergalactic magnetic field (IGMF) and extragalactic background light (EBL) using 
$\gamma\gamma$-interactions via pair production. Despite many optical spectroscopic campaigns for measuring the redshift of 
blazars, only a small fraction of BL Lacs have well known redshifts \cite{Piranmonte2007,Shaw2013,Paiano2017}. 
However, redshift measurements of a large fraction of FSRQs using spectroscopic observations have been reported \cite{Shaw2012}. 
A new physical method for the measurement of redshift of AGN using the time lag between light curves in high and low energy bands is 
also proposed \cite{Yoshi2014}. This method is based on the quantitative model of the dust reverberation in AGN which relates 
the absolute luminosity of the source with the time lag but it is useful for sources at smaller redshifts only. 
Prandini et al. (2010) have pioneered a method to constrain the redshift of blazars using combined GeV-TeV 
$\gamma$--ray observations \cite{Prandini2010}. This method strongly depends on the model for the density of EBL photons in the 
Universe which is not exactly known until today \cite{Dominguez2013,Singh2014,Franceschini2017,Fermi2018} and use of different 
EBL models which may lead to large uncertainty  in the derived redshift for a given source. Another method proposed by Qin et al. (2018) 
estimates redshift of three BL Lacs through the fitting of their broad-band SED using a single zone leptonic model \cite{Qin2018}. 
This method also depends on the EBL model and the model for multi-wavelength emission from blazars which are not universally applicable 
to all sources.

In this work, we propose an artificial intelligence based approach using Artificial Neural Network (ANN) to constrain the redshift 
of blazars from the GeV-TeV $\gamma$--ray observations. This method is based on the assumption that the observed spectral-break 
between MeV-GeV spectra from the \emph{Fermi}-Large Area Telescope (LAT) and TeV spectra from the ground-based observations of blazars 
strongly depends on the redshift of the source. Such a correlation could arise, for example, because of attenuation of TeV photons by 
EBL during their propagation towards Earth. The paper is structured as following: in Section 2, we discuss the $\gamma$--ray 
spectra of blazars. The blazar sample used in this work is described in Section 3. In Section 4, a brief description of the ANN 
methodology is presented. The results are discussed in Section 5. Finally, we have summarized the study in Section 6.

%-------------------------------Table-1-Blazar-list-for training-------------------------------------------------
\begin{table}
\begin{center}
\caption{List of blazars selected for ANN training from the TeGeV/TeV catalog with known redshift.}
\vspace{1.0cm}
\begin{tabular}{|c|c|c|}
\hline
Name  		&Type	  &Redshift (z) \\
\hline
PKS 2005-489    &HSP       &0.071  \\ 	           
H 2356-309      &HSP       &0.165   \\	           
Mrk 180         &HSP       &0.045   \\ 	           
PKS 0548-322    &HSP       &0.069   \\	           
1ES 1011+496    &HSP       &0.212   \\	           
RGB J0152+017   &HSP       &0.08    \\	           
1ES 0806+524    &HSP       &0.138   \\	           
RGB J0710+591   &HSP       &0.125   \\	           
RBS 0413        &HSP       &0.19    \\	           
1H 0323+022     &HSP       &0.147   \\	           
VER 0648+152    &HSP       &0.179   \\	           
B3 2247+381     &HSP       &0.1187  \\	           
1RXS J1010-311  &HSP       &0.1426  \\	           
1ES 1727+502    &HSP       &0.055   \\	           
1ES 0120+340    &HSP       &0.272  \\	           
Mrk 421         &HSP       &0.031   \\	           
1ES 1741+196    &HSP       &0.084   \\	           
H 1426+428      &HSP       &0.129   \\	           
1ES 1959+650    &HSP       &0.048   \\	           
PKS 2155-304    &HSP       &0.116   \\	           
1ES 1218+304    &HSP       &0.182   \\	           
1ES 1101-232    &HSP       &0.186   \\
1ES 0033+595    &HSP       &0.086   \\
TXS 1055+567    &HSP       &0.143   \\	           
1H 0658+595     &HSP       &0.125   \\			           
PKS 0301-243	&HSP       &0.266    \\	
1H 1013+498     &HSP       &0.212  \\
3C 66A          &ISP       &0.444   \\
VER J0521+211   &ISP       &0.108   \\	           
S5 0716+714     &ISP       &0.31    \\
1ES 2202+420    &ISP       &0.069   \\
AP Librae       &LSP       &0.049   \\
3C 279          &LSP       &0.5362  \\	           
PKS 1222+21      &LSP       &0.432   \\		  
PKS 1510-089    &LSP       &0.361   \\
\hline
\end{tabular}
\label{tab:table1}
\end{center}
\end{table}
%----------------------------------------Section-2-Gamma-ray Emission-----------------------------------------------------------------------
\section{GeV-TeV $\gamma$--ray Spectra of Blazars}
The origin of $\gamma$--ray emission from blazars has not been completely understood  so far and it is attributed to the inverse 
Compton (IC) scattering of low energy photons by the relativistic electrons in the leptonic scenario and to the synchrotron radiation 
of ultrarelativistic protons in the hadronic models \cite{Romero2012}. However, the $\gamma$--ray observations of blazars suggest that 
the observed $\gamma$--ray emission from most of the blazars can be described by a power law in a given energy range. Therefore, the 
differential spectrum of $\gamma$--ray photons observed from a blazar by an instrument can be expressed as 
\begin{equation}
	F_{obs}(E) ~ \propto~ E^{-\Gamma}
\end{equation}
where $\Gamma$ is the power law spectral index. The \emph{Fermi}-Large Area Telescope (LAT) provides an excellent measurement of $\Gamma$ 
in the GeV energy band where the peak of the HE component in the SED is observed for most of the blazars  \cite{Atwood2009}. 
For $\Gamma \le$ 2, the \emph{Fermi}-LAT detects the photons with energy less than the peak energy of the HE component in the braod-band SED, 
which is generally referred to as the hard spectrum. The observations with the ground-based atmospheric Cherenkov telescopes measure the 
spectral index of TeV photons in the VHE regime, which belongs to the soft or steep portion of the spectrum. Therefore, the observations 
from the \emph{Fermi}-LAT overlapping with the ground-based observations can provide a good shape of the $\gamma$--ray spectra of blazars 
in the GeV-TeV energy range and also the peak of HE component in SED.
%----------------------------Table 2 ANN Testing ------------------------------------------------------
\begin{table}
\begin{center}
\caption{Summary of the blazar sample from the TeGeV/TeV catalog used for testing ANN.}
\vspace{1.0cm}
\begin{tabular}{|c|c|c|c|c|}
\hline
Name  		&Type	  &z(Known)  &z(ANN)	&Uncertainty in z ($\%$)\\
\hline
1ES 0502+675    &HSP       &0.341    &0.358     &4.98\\	   
Mrk 501          &HSP       &0.034    &0.040     &17.60\\ 
1ES 0647+250    &HSP       &0.203    &0.206     &1.47\\
1ES 1440+122    &HSP       &0.162    &0.156     &3.70\\
1ES 2344+514    &HSP       &0.044    &0.049     &11.36\\
PKS 0447-439    &HSP       &0.343    &0.329     &4.08\\
1ES 1312-423    &HSP       &0.105    &0.107     &1.90\\
PG 1553+113     &HSP       &0.129    &0.131     &1.55\\ 
B2 1219+28      &ISP       &0.102    &0.103     &0.98\\
1ES 1215+303    &LSP       &0.130    &0.110     &15.38\\
\hline
\end{tabular}
\label{tab:table2}
\end{center}
\end{table}
%----------------------------------------Section-3-Sample-------------------------------------------------------------------
\section{The Blazar Sample}
\begin{figure}
\begin{center}
\includegraphics[scale=0.62,angle=-90]{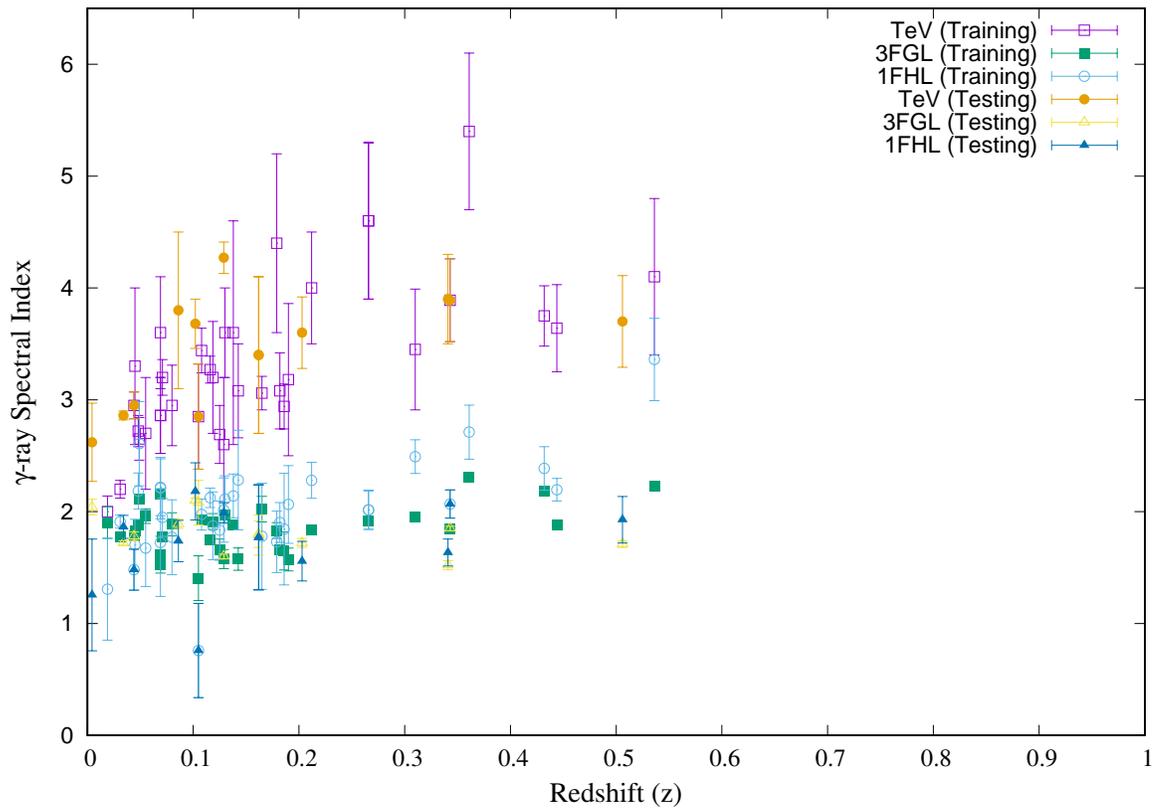}
\caption{$\gamma$--ray spectral indices as a function of redshift for the sample of blazars selected for ANN training and testing.} 
\label{fig:fig1}
\end{center}
\end{figure}

The launch of Large Area Telescope (LAT) onboard the \emph{Fermi} satellite in 2008 has opened a new window to 
explore the $\gamma$--ray sky using space-based observations in the GeV energy band \cite{Atwood2009}. The \emph{Fermi}-LAT 
observations combined with the ground-based Cherenkov telescopes operating in sub-TeV energy range provide an unique 
opportunity to obtain a number of innovative scientific results in astrophysics and cosmology. Based on the first 
four years of data taken during August 2008 to July 2012, more than 1100 blazars have been reported in the third 
\emph{Fermi}-LAT source catalog (3FGL) in the energy range 0.1-300 GeV \cite{Acero2015}. The first \emph{Fermi}-LAT 
catalog of high energy sources above 10 GeV (1FHL) describes more than 300 blazars detected in the energy range 
10 GeV--500 GeV using first three years of data from August 2008 to August 2011 \cite{Ackermann2013}. 
The $\gamma$--ray emission from the majority of the blazars reported in the 3FGL and 1FHL catalogs is described by a power law 
in the energy bands 0.1-300 GeV and 10-500 GeV respectively. The combined observations from the 3FGL and 1FHL catalogs 
provide an important data set to determine the HE $\gamma$--ray spectra of blazars in the wide energy range of 0.1-500 GeV.
However, more than 50$\%$ of the blazars reported in these catalogs lack their redshift measurements. The online TeGeV 
catalog \cite{Carosi2015} and TeV catalog\footnote{http://tevcat.uchicago.edu/} provide results from the VHE $\gamma$--ray 
observations of the sources by the past and current generation of ground-based Cherenkov telescope. More than 70 blazars 
have been discovered at TeV energies in the extragalactic Universe by the ground-based Cherenkov telescopes. An interactive 
version of the above three catalogs (3FGL, 1FHL and TeGeV) is publicly available at Space Science Data Center 
(SSDC)\footnote{https://fermi.ssdc.asi.it/}. This provides a very useful quasi-simultaneous data base of blazars in 
three $\gamma$--ray energy bands. The redshift measurements of these sources using optical observations suggest that most 
of the TeV blazars are located at redshift z $<$ 1. BL Lac blazars are observed at low redshifts while FSRQs have relatively 
higher redshift \cite{Giommi2012}. In the present work, we have selected a sample of 45 blazars from the TeGeV or TeV catalog with 
their $\gamma$--ray spectral measurements available in 3FGL and 1FHL catalogs. The $\gamma$--ray spectra of blazars selected in 
the sample are described by a power law with spectral indices $\Gamma_{TeV}$, $\Gamma_{3FGL}$ and $\Gamma_{1FHL}$ in the TeV, 3FGL 
and 1FHL catalogs respectively. We have randomly classified the blazar sample in two groups: training and testing data sets. 
A large fraction of the blazars with known redshift from this sample is used for training the ANN methodology based on the 
spectral measurements of these source. The list of TeV blazars ($\sim$ 35) used for training the ANN is given in 
Table \ref{tab:table1}. About 10 blazars with well known redshifts are used as testing data set to validate the 
procedure for predicting the unknown redshift using their $\gamma$--ray spectral measurements from \emph{Fermi}-LAT and 
TeV instruments. The blazar sample used for testing ANN is summarized in Table \ref{tab:table2}. It is important to note  
that few blazars in the sample have more than one redshift measurements in the literature. For such blazars, we have 
used only those redshift values which are commonly reported in various $\gamma$--ray catalogs in this study. Also, 
the spectral properties of blazars detected by the \emph{Fermi}-LAT suggest that the distribution of power law 
spectral index of photons in the energy range above 100 MeV is strongly correlated with the blazar types \cite{Abdo2010b}.
A departure from the single power law photon spectrum is mainly observed for ISP and LSP types of blazars whereas this 
feature is absent in the HSP class of blazars. Therefore, we have selected only those \emph{Fermi}-LAT blazars in the sample 
which are described by a single power law in MeV-GeV and TeV energy bands. The distribution of $\gamma$--ray spectral indices 
($\Gamma_{TeV}$, $\Gamma_{3FGL}$ and $\Gamma_{1FHL}$) of the blazar sample as a function of their known redshift (z) is shown in 
Figure \ref{fig:fig1}. It is evident from Figure \ref{fig:fig1} that most of the blazars in the sample are populated up to the 
redshift z$<$0.3 and the measured $\gamma$--ray spectral indices from the TeV instruments ($\Gamma_{TeV}$) show a definite 
correlation with the redshift. 

%---------------------------Figure:1 Spectral Index---------------------------------------------------------------------

%-------------------------------------Section-4:ANN---------------------------------------------------------------
\section{Artificial Neural Networks}
The artificial neural network (ANN) is a mathematical construct for data prediction by recognising the  correlations and patterns 
in the input training data sets. The ANN based computational techniques efficiently replicate the behavior of human brain. 
ANN is collection of interconnected processing units known as nodes similar to the human brain which consists of biological cells 
called neurons. The strength of interconnection among neurons is characterized by weights. A neuron in the network generates a single 
output from multiple inputs. The ANN system has at least three layers namely input, output and hidden with different number of 
neurons. More than one hidden layer can also be used depending on the complexity of the problem. The hidden layer links input and 
output layers in a very complex way. In the simplest form, the data is supplied to the input layer neurons which generally acts as 
a buffer and passes the data to the hidden layer. The hidden layer produces an output using the non-linear transformations of the 
signal from the input layer and passes the data to the output layer to predict the output. The output generated by the ANN is 
compared with the desired output to estimate the error \cite{Sinkus1995}. The ANN method learns by adjusting the weights such that 
the produced error is reduced until the final output is computed \citep{Dhar2010a}. The number of neurons in a network is optimized from 
the nature of the problem at hand. The performance of an ANN as a whole is optimized using input training data sets with correct output 
for a given application. The first technique using artificial neurons was developed in 1943 to perform the logical operations 
\cite{ANN1943}. Nowadays ANN based computational techniques have received ample applications in astrophysics, science and other
diverse areas.

\subsection{Training of ANN: Levenberg-Marquardt Method}
The training of ANN includes the search for appropriate values of network parameters by minimizing the difference between the 
predicted and correct values also known as the error function. A feed-forward method is the most commonly used ANN algorithm. 
However, varied algorithms exist under the ANN domain depending upon the choice of error function \cite{Dhar2013}. The learning 
schemes followed by Back-propagation based on the gradient-descent methods have several limitations \cite{Rumelhart1986}. 
In the gradient-based algorithms, it is difficult to obtain a unique set of optimal parameters due to the existence of multiple 
local minima. On the other hand, the Levenberg-Marquardt method is a compromise between the Newton 
method and the gradient descent method employed by the backpropagation algorithm. The advantage of this coalition is that while 
the Newton method converges very rapidly near a local or the global minimum but may also diverge, the gradient descent is assured of 
convergence through a proper selection of the step size, but converges slowly. For example, consider the optimization of a second order 
function F(${\bf w}$) and let ${\bf g}$ and ${\bf H}$ be the gradient vector and the Hessian respectively. According 
to the Levenberg-Marquardt method, the optimum adjustment $\Delta{\bf w}$ applied to the parameter ${\bf w}$ is defined as
\begin{equation}\label{weight}
	\Delta{\bf w} = \left[{\bf H} + \Lambda {\bf I}\right]^{-1} {\bf g}
\end{equation}
where ${\bf I}$ is the identity matrix of the same dimension as ${\bf H}$ and ${\bf \Lambda}$ is a regularizing, loading/blending 
parameter that forces the sum matrix $\left[{\bf H}+ \Lambda{\bf I}\right]$ to be positive definite and well conditioned throughout 
the computation. Now considering the application to the present work which has a single output redshift (z), the network is trained 
by minimizing the cost or the error function 
\begin{equation}\label{errf}
	\Omega_{av}({\bf w}) = \frac{1}{2N} \sum^{N}_{i=1}\left[d(i) -F\left({\bf x}(i), {\bf w}\right)\right]^2
\end{equation}
where (${\bf x}(i), d(i)$) is the training sample comprising of spectral indices $\Gamma_{TeV}$, $\Gamma_{3FGL}$, $\Gamma_{1FHL}$ 
and redshift (z). F$\left({\bf x}(i), {\bf w}\right)$ is the approximating function realized by the network. The synaptic weights of 
the network are arranged in some orderly manner to form the weight vector ${\bf w}$. The gradient and the Hessian of the error 
function $\Omega_{av}({\bf w})$ to be minimized are respectively defined as
\begin{eqnarray}\label{grad}
	{\bf g}(w)= \frac{\partial{\Omega_{av}({\bf w})}}{\partial{\bf w}}
                  = -\frac{1}{N} \sum^{N}_{i=1} \left[d(i)-F({\bf x}(i), {\bf w})\right]\frac{\partial F({\bf x}(i),
                   {\bf w})}{\partial{\bf w}}
\end{eqnarray}
and 
\begin{eqnarray}
	{\bf H}({\bf w})= \frac{\partial^2{\Omega_{av}({\bf w})}}{\partial {{\bf w}^2}}      
                        = \frac{1}{N}\sum^{N}_{i=1} \left[\frac{\partial F(\textbf{x}(i), \textbf{w})}{\partial \textbf{w}}\right]
                          \left[\frac{\partial F(\textbf{x}(i), \textbf{w})}{\partial\textbf{w}}\right]^T  \\
                        = -\frac{1}{N} \sum^{N}_{i=1} \left[d(i)-F(\textbf{x}(i), \textbf{w})\right]\frac{\partial^2
                           {F(\textbf{x}(i), \textbf{w})}}{\partial{\textbf{w}^2}}              
\end{eqnarray}  
Substituting the solutions obtained for above equations in the equation \ref{weight}, the desired weight adjustment 
$\Delta{\bf w}$ is computed for each iteration of the Levenberg--Marquardt method. However from a practical perspective, 
the computational complexity for calculating ${\bf H}({\bf w})$ is demanding especially when the dimentionality of the 
weight ${\bf w}$ is high. This computational complexity is due to the complex nature of the Hessian ${\bf H}({\bf w})$. 
Fortunately this difficulty is mitigated by approximating the Hessian simply as 
\begin{equation}\label{hess}
	{\bf H}({\bf w}) \approx \frac{1}{N}\sum^{N}_{i=1} \left[\frac{\partial F(\textbf{x}(i), \textbf{w})}
                                {\partial\textbf{w}}\right] \left[\frac{\partial F(\textbf{x}(i), \textbf{w})}
                                {\partial\textbf{w}}\right]^T
\end{equation}
This approximation method is recognized as the outer product of the partial derivative $\frac{\partial{{\bf F}}({\bf w},
{\bf x}(i))}{\partial {\bf w}}$ with itself, averaged over the training sample. Clearly, the approximate version of the 
Levenberg-Marquardt algorithm based on the gradient vector (equation \ref{grad}) and the Hessian (equation \ref{hess}) is 
a first order method of optimization which is well suited for non-linear least square estimation problems. 
\par
The loading/blending factor ${\bf \Lambda}$ plays a very crucial role in the implementation of Levenberg--Marquardt algorithm. 
If the parameter ${\bf \Lambda}$ is set to 0 , then equation \ref{weight} reduces to the well known traditional Newton method. 
However if we assign a large value to  ${\bf \Lambda}$ such that ${\bf \Lambda}{\bf I} $ becomes more important in 
equation \ref{weight} compared to Hessian ${\bf H}$, then the Levenberg--Marquardt algorithm functions effectively like 
the gradient descent method employed in the tradionally used backpropagation algorithm. From this, we conclude that 
at each iteration of the algorithm, the value assigned to ${\bf \Lambda}$ should be just large enough to maintain the sum 
matrix ($\textbf{H} + {\bf \Lambda} {\bf I}$) in a positive definite form. The specific formulation for selection of the 
parameter ${\bf \Lambda}$ has been proposed by \cite{Press1988} as follows:
\begin{itemize}
	\item Compute $\Omega_{av}(\textbf{w})$ at iteration (n-1)
	\item Choose a modest value for ${\bf \Lambda}$, say ${\bf \Lambda} =10^{-3}$
	\item Solve equation \ref{weight} for the adjustment of $\Delta {\bf w}$ at iteration (n) and evaluate 
             $\Omega_{av}({\bf w} + \Delta{\bf w})$
	\item If $\Omega_{av}({\bf w} + \Delta{\bf w}) \geq \Omega_{av}({\bf w})$ increase ${\bf \Lambda}$ by factor 10 or more
              go back to the above step.
	\item If $\Omega_{av}({\bf w} + \Delta{\bf w}) \leq \Omega_{av}({\bf w})$, decrease ${\bf \Lambda}$ by factor 
              10, update the trial solution ${\bf w} \rightarrow {\bf w} +\Delta{\bf w}$ and go back to step 3.
\end{itemize}
An indepth comparison of the popular back-propagation (generally used in the ANN applications) and Levenberg--Marquardt method 
(used in the present work) has been studied in detail and the superior performance of the latter method is also 
demontrated \cite{Dhar2010b}. Thus, we have decided to use more efficient Levenberg--Marquardt method in the present study.
\par
A properly trained ANN can be thought of as an expert in the category of information it has 
been given to analyze. Thus ANN is a massively parallel distributed processor made up of simple processing units that has a 
natural propensity for storing experimental knowledge and making it available for use. It is similar to the human brain functioning 
in the sense that the knowledge is acquired by the network from its environment through a learning process and interneuron 
connection strenghts, known as synaptic weights are used to store the acquired knowledge. The procedure used to acquire 
the knowledge is called the learning algorithm, the function of which is to modify the synaptic weights of the network 
in an orderly fashion to attain the desired objective based on some well established error minimizing methods. 
The modification of the synaptic weights provides the traditional method for the design of the neural networks. Such an approach 
is closest to linear adaptive filter theory, which is well established and applied in many diverse fields.
%-----------------------------------Table-3--ANN--------------------------------------------
\begin{table}
\begin{center}
\caption{Mean Square Error (MSE) for different number of neurons used in the training of ANN.}
\vspace{1.0cm}
\begin{tabular}{lclll}
\hline
Neurons 	& MSE 			       &$\sigma$(Training)	 &$\sigma$(Testing) \\         
\hline
3 		& 3.08$\times$10$^{-3}$ 	&48.7$\%$ 	 &54.3$\%$ \\
4 		& 1.05$\times$10$^{-3}$ 	&17.1$\%$ 	 &19.1$\%$ \\
5 		& 1.03$\times$10$^{-3}$ 	&16.6$\%$ 	 &48.2$\%$ \\
10 		& 2.02$\times$10$^{-6}$         &1.97$\%$ 	 &75.3$\%$ \\
\hline
\end{tabular}
\label{tab:table3}
\end{center}
\end{table}

\subsection{Application of ANN}
\begin{figure}
\begin{center}
\includegraphics[scale=0.62,angle=-90]{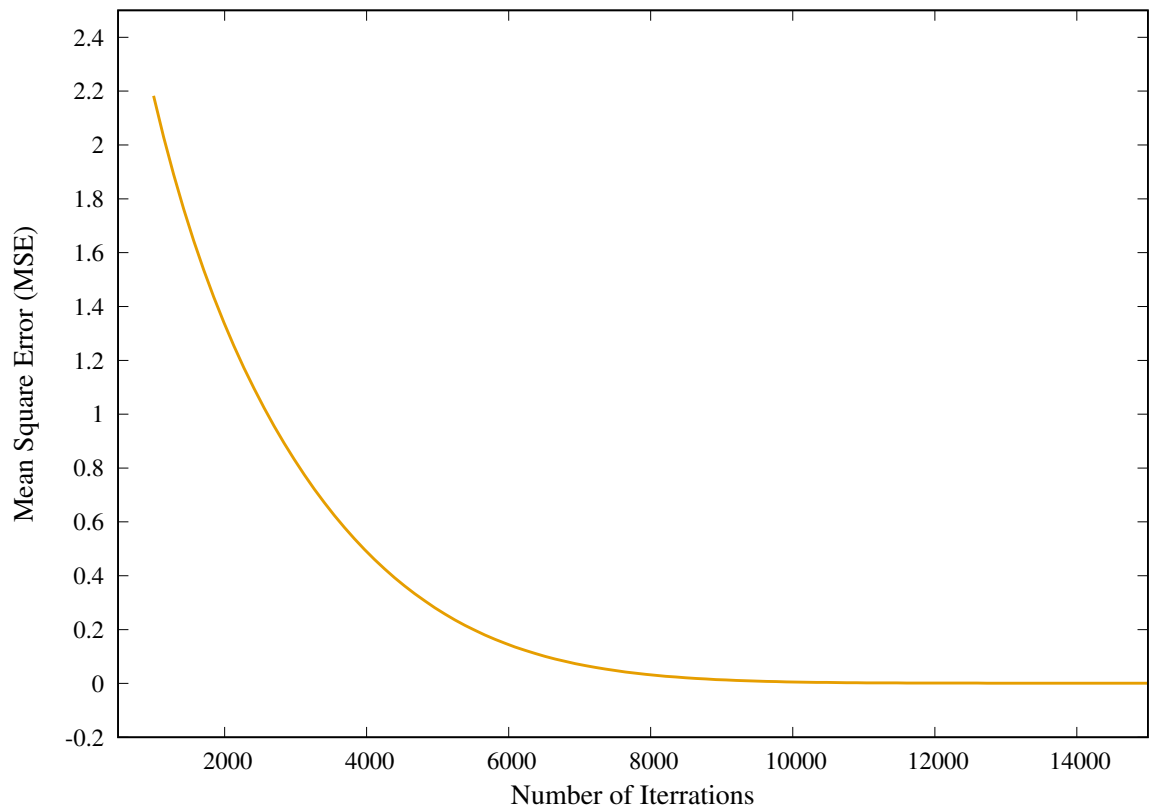}
\caption{Optimization of the number of iterations using mean square error (MSE) estimation for Levenberg–Marquardt algorithm with 4 neurons.} 
\label{fig:fig2}
\end{center}
\end{figure}

In the present work, we use ANN to constrain the redshift of a sample of blazars using the combined $\gamma$--ray spectral 
measurements in GeV-TeV energy bands from \emph{Fermi}-LAT and ground-based Cherenkov telescopes. The main objective is to 
predict the redshift (z) of a blazar using the set of measured spectral indices $\Gamma_{TeV}$, $\Gamma_{3FGL}$ and $\Gamma_{1FHL}$.
Therefore, the data set for training ANN in this work consists of 3 inputs: $\Gamma_{TeV}$, $\Gamma_{3FGL}$ and $\Gamma_{1FHL}$ 
and 1 output which is the redshift (z) of the sources to be predicted. We have used $\sim$ 35 blazars (listed in Table \ref{tab:table1}) 
for training the ANN. The first step towards using the Levenberg-Marquardt method, is to find the optimized number of neurons which are 
employed for training. A trial and error method is employed for selecting the optimized number of neurons. We started with 3 neurons 
in one hidden layer and changed it to 4, 5, and 10 and observed the performance of the ANN. The goodness of the performance is 
determined by the Mean Square Error (MSE) in each case. The MSE is defined as 
\begin{equation}
	MSE = \frac{1}{N} \Sigma_{i=1}^{N} \left(Y_{true}^i - Y_{pred}^i\right)^2
\end{equation}
where $N$ is the number of samples, $Y_{true}^i$ and $Y_{pred}^i$ are the true and predicted (by ANN) values of the expected 
output for $i$-th example. This function is minimized during the training process for the ANN. Variation of number of neurons 
versus MSE is given in Table \ref{tab:table3}. It is clear from the table that the lowest MSE of 2.02$\times$10$^{-6}$ is obtained 
for 10 neurons. Levenberg-Marquardt algorithm being inherently a very powerful method, one has to be extremely careful to avoid any 
overfitting while choosing the number of neurons in the hidden layer. Overfitting of data can result due to choosing a higher number 
of neurons than required. This leads to the memorising 
of the training data rather than generalizing it. A simple method to check the overfitting is to check the results on test data which 
have not been seen by the network during training. The blazars used for testing from the sample are summarized in Table \ref{tab:table2}. 
For a good generalization the performance on test and training data should be nearly similar (slightly better results are expected for 
training data). The performance of ANNs with different number of neurons on the training and testing data is compared in 
Table \ref{tab:table3}, where columns 3 and 4 are the standard deviations ($\sigma$) corresponding to the training and testing data 
sets respectively. It is obvious from the table that overfitting starts after 5 neurons due 
to large discrepency between the values of standard deviation for training and the testing. In view of this, there is no need to increase the 
number of neurons further as that would only have worsened the results. Best results were obtained with 4 neurons (row 2 in Table \ref{tab:table3}) 
where close matching is found between the training and the testing results. Thus 4 neurons in one hidden layer can be employed for the prediction 
of unknown redshifts of blazars. Next, we have to optimize the number of iterations, which happens to be a slightly easier task. Once 4 neurons are 
chosen, we have optimized the number of iterations using the MSE. The variation of the MSE as a function of number of iterations with 4 neurons 
is shown in Figure \ref{fig:fig2}. We observe that the MSE does not vary after $\sim$ 11000 iterations. This suggests that approximately 15000 iterations 
performed for the error reduction are sufficient for the problem at hand.  

%---------------------------------------------Figure:2-Mean Square Error------------------------

%---------------------------------------Section-5:Result and Discussion-----------------------------------------------
\section{Results and Discussion}
\begin{figure}
\begin{center}
\includegraphics[scale=0.62,angle=-90]{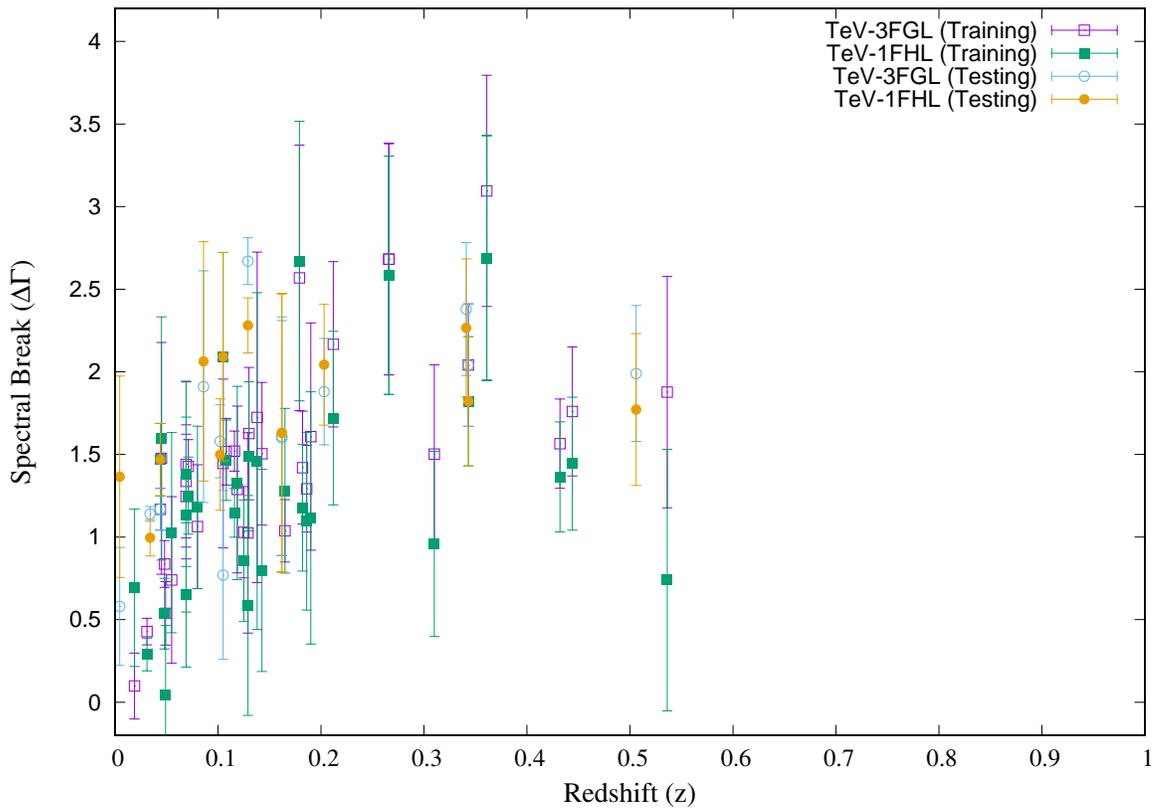}
\caption{Spectral break between the TeV and MeV-GeV $\gamma$--ray spectra as a function of redshift for the blazar sample 
	 used in the ANN training and testing.} 
\label{fig:fig3}
\end{center}
\end{figure}

In this work, we used a sample of approximately 45 blazars which have been detected at GeV energies by the \emph{Fermi}-LAT 
and at TeV energies by the ground-based $\gamma$--ray instruments for constraining their redshifts using ANN. A close look of the 
test data sample listed in Table \ref{tab:table2} (columns 3 and 4) suggests that the known redshifts of the blazars (z) are in good 
agreement with the values predicted by the Levenberg - Marquardt based artificial intelligence using ANNs. The results are found to be 
consistent well within an uncertainty of $\sim$ 18$\%$ as listed under column 5 in Table \ref{tab:table2}. This indicates that the 
$\gamma$--ray spectral indices of blazars obtained from the \emph{Fermi}-LAT and ground-based VHE observations can be used to  
constrain or predict their redshifts with  good accuracy. The $\gamma$--ray spectral indices shown in Figure \ref{fig:fig1} suggests 
that the \emph{Fermi}-LAT spectral indices reoprted in 3FGL and 1FHL catalogs are nearly similar. However, the TeV spectral indices 
($\Gamma_{TeV}$) show siginificant softening with increasing redshift as compared to $\Gamma_{3FGL}$ and $\Gamma_{1FHL}$. This implies 
a clear spectral break in the $\gamma$--ray spectra of blazars observed with  the \emph{Fermi}-LAT and TeV instruments. The observed 
$\gamma$--ray spectral break ($\Delta \Gamma$), which is defined as the difference between the measured GeV and TeV spectral indices, 
is shown in Figure \ref{fig:fig3} as a function of redshift (z). It is observed from Figure \ref{fig:fig3} that the spectral break is 
non-zero for all the blazars used in the sample for training and testing the ANN. For the blazar sample used in training the ANN, 
the observed spectral break is a linear function of the redshift (z), which can be mathematically expressed as 
\begin{equation}
	\Delta \Gamma = \Gamma_{TeV} - \Gamma_{LAT} = az + b
\end{equation}	
where $a$ and $b$ are the parameters, and $\Gamma_{LAT}$ is the spectral index measured by \emph{Fermi}-LAT ($\Gamma_{3FGL}$ or 
$\Gamma_{1FHL}$). For the spectral break between $\Gamma_{TeV}$ and $\Gamma_{3FGL}$, the best fit parameter values are found 
to be $a = 3.78\pm0.74$ and $b = 0.67\pm0.09$. Similarly, for the spectral break between $\Gamma_{TeV}$ and $\Gamma_{1FHL}$,  
we get  $a = 3.04\pm0.77$ and $b = 0.60\pm0.11$. This indicates that the values of parameters $a$ and $b$ are similar 
within error bars for both the cases. Therefore, the GeV spectral indices in the energy range 0.1-500 GeV and TeV spectral indices 
above $\sim$ 100 GeV of the blazars lead to a redshift dependent spectral break in their $\gamma$--ray spectra. The origin of the 
observed spectral break in the $\gamma$--ray spectra of blazars is not clearly known and is attributed to several factors like an 
intrinsic break in the emitted spectrum of the source (intrinsic curvature) or softening of the TeV spectrum due to EBL absorption 
(extrinsic curvature) in the extragalactic space \cite{Stecker1992,Stecker2006}. The attenuation of TeV $\gamma$--ray photons during the 
propagation from source to Earth via $\gamma_{TeV}\gamma_{EBL}\rightarrow e^-e^+$ strongly depends on the local density of EBL photons and 
redshift of the source. Therefore, any uncertainty in the EBL photon density will lead to the incorrect estimation of the spectral break as 
investigated by \cite{Prandini2010}. Also the intrinsic curvature in the emitted spectrum depends on the particle distribution considered in 
a given blazar emission model \cite{Qin2018}. However, the artificial intelligence methodology used in the present work is entirely based 
on the observed spectral break in GeV-TeV spectra of blazars and does not consider any physical effect for the curvature. The observed 
$\gamma$--ray spectral break is found to be a distinct property of blazars but its redshift evolution has not been properly explained by 
the blazar physics. It can provide important information about the location and structure of the $\gamma$--ray emission region in the blazar jet.
%-------------------------------------Figure:3-Spectral break----------------------------------------

%----------------------------------------Section:6-Summary------------------------------------------------
\section{Summary}
We have demonstrated that the spectral break observed in the GeV-TeV $\gamma$--ray spectra of blazars can be effectively utilised 
to constrain their redshift by using an ANN based artificial intelligence technique. We find that an ANN with 4 neurons in 
the hidden layer and with 15000 iterations can perform very well for constraing the redshift of blazars using the Levenberg-Marquardt 
algorithm. This technique only depends on the observational features of the  $\gamma$--ray spectra of blazars in different energy bands. 
The redshift of blazars can be constrained or predicted by this technique with an uncertainty of about 18$\%$ using only the 
$\gamma$--ray spectral information from the \emph{Fermi}-LAT and ground-based observations. The availability of more blazars in 
the training sample of ANN will reduce the uncertainty in the predicted values. Therefore, the results from the upcoming CTA 
(Cherenkov Telescope Array) observatory will be very useful for such applications based on artificial intelligence approach. 
	      
%-------------------------------------Acknowledgements----------------------------------------------
\section*{Acknowledgments}
Part of this work is based on archival data, software or online services provided by the Space Science Data Center -ASI.

%-----------------------------------------References--------------------------------------------------------

\end{document}